\documentclass[a4paper]{jpconf}
\usepackage{graphicx}
\begin{document}
\title{The Fano-Rashba effect}

\author{Lloren\c{c} Serra$^{1,2}$ and David S\'anchez$^1$}

\address{
$^1$ Departament de F\'{\i}sica, Universitat de les Illes Balears,
E-07122  Palma de Mallorca, Spain\\*
$^2$ Institut Mediterrani d'Estudis Avan\c{c}ats IMEDEA (CSIC-UIB), 
E-07122 Palma de Mallorca, Spain}

\ead{llorens.serra@uib.es, david.sanchez@uib.es}

\begin{abstract}
We analyze the linear conductance of a semiconductor quantum wire containing
a region where a local Rashba spin-orbit interaction is present. 
We show that Fano
lineshapes appear in the conductance 
due to the formation of quasi bound states which interfere
with the direct transmission along the wire, a mechanism that we term 
the Fano-Rashba effect. 
We obtain the numerical solution of
the full Schr\"odinger equation using the quantum-transmitting-boundary method.
The theoretical analysis is performed using the coupled-channel model, finding
an analytical solution by ansatz. The complete numerical 
solution of the coupled-channel equations is also discussed, showing the validity 
of the ansatz approach.
\end{abstract}

\section{Introduction}

The spin-orbit (SO) interactions in semiconductor nanostructures are currently
attracting much interest due to their central role in the characterization
and control of spin properties of these systems \cite{zu04}. 
Particular attention is devoted to the Rashba
interaction in two-dimensional quantum wells ---a SO coupling
stemming from the well inversion asymmetry in the growth direction
whose intensity can be manipulated by
electrical gating \cite{ras60,nit97}. The Rashba interaction is also the 
underlying physical 
mechanism for the operation of the proposed spin transistor, one of the 
paradigms of spintronic device, exploiting the electron's spin to control
the charge current between two ferromagnetic contacts \cite{dat90}. 

In a quantum wire with {\em uniform} SO interaction the energy subbands 
deviate from the simple parabolic behavior, with band 
splittings and anticrossings that yield a modified 
wire conductance and the possibility of spin-textured states \cite{scha04}.
In this
work we consider a wire with {\em nonuniform} Rashba interaction,
localized to a finite region acting as a spin scattering center
(a Rashba dot) that affects the wire conductance properties.  It will 
be shown with numerical and analytical results that this 
geometry leads in a natural way to the appearance of Fano resonance 
profiles in the energy dependence of the conductance that depend
quite sensitively on the properties of the Rashba dot, such as 
its dimensions and SO intensity. We refer to this influence of the 
Rashba SO coupling on the wire conductance as the 
Fano-Rashba effect.

Fano resonances \cite{fano} are a general phenomenon that has been observed in 
different fields, such as atomic physics \cite{ada49}, Raman scattering \cite{cer73} 
and mesoscopic electron transport \cite{gor00}.
Fano-resonance physics appears 
wherever there is an interference between two paths, one corresponding 
to direct transmission and the other to the passage through a quasi-bound state
lying nearby in energy. 
As a consequence, characteristic
asymmetric lineshapes appear in the most general case, with conductance
dips in which transmission is greatly quenched. Indeed, for the 
case of scattering centers that may be modelled as attractive potentials 
it was shown that an exact transmission zero should 
always be present \cite{gur93,noc94}.
In our case of a Rashba dot the existence of transmission zeros is not always 
guaranteed and only for some specific dimensions and intensities of SO 
coupling the wire conductance vanishes at the dip position.

A useful theoretical framework to understand Fano resonances is the 
coupled-channel model (CCM). In this approach the wave function is 
approximated by the superposition of different channel components, each
of them obeying an equation containing the channel background
contribution and also the coupling with the other channels. We shall 
compare the prediction of CCM with the exact numerical solution obtained 
from a grid discretization of the Schr\"odinger equation. The prediction
of the CCM will be analyzed using the ansatz for the 
quasi-bound state channel proposed by N\"ockel and Stone \cite{noc94}, 
and also with a 
full numerical solution of the CCM equations. In this way we shall
explicitly proof the validity of the ansatz. This analysis  
is an extension of our recent work \cite{san06} and is also related to the 
study in Ref.\ \cite{zhan05} where numerical results for the transmission of 
wires with a modulated Rashba intensity were obtained.
A localized Rashba interaction in a quantum wire was also considered
in Ref.\ \cite{egu02} for the study of shot noise and entanglement 
within a beam splitter configuration.

\section{The physical system}

\subsection{Hamiltonian}

We assume the effective-mass model for the conduction band states
of a two-dimensional electron gas in GaAs and consider a
parabolic confinement in the $y$ direction, yielding a ballistic 
quantum wire
along $x$. The system Hamiltonian reads
\begin{equation}
\label{eq1}
H={p_x^2+p_y^2\over 2m}+\frac{1}{2}m\omega_0^2y^2+H_R\; .
\end{equation}
The oscillator is used to define our energy $\hbar\omega_0$
and length $\ell_0=\sqrt{\hbar/m\omega_0}$ units.
The Rashba Hamiltonian $H_R$ that we consider in Eq.\ (\ref{eq1}) is characterized by 
an intensity $\alpha(x)$
vanishing everywhere except for $0<x<\ell$ where it takes the value
$\alpha_0$.\footnote[1]{
In the numerical applications we take $\alpha(x)=\alpha_0 [f(x-\ell)-f(x)]$, with 
$f(x)=1/(1+e^{x/\sigma})$ and $\sigma$ a small diffusivity. 
The results are not very sensitive 
to the precise value of $\sigma$
provided $\sigma<\ell$.}
In detail, $H_R$ reads
\begin{equation}
\label{eq2}
H_R= \frac{\alpha(x)}{2\hbar}  \left( p_y\sigma_x-p_x\sigma_y \right)
+ {\it h.c.} \; ,
\end{equation}
where $\sigma_x$ and $\sigma_y$ are the Pauli matrices and the Hermitian
conjugation is used to ensure Hermiticity. The inset to Fig.\ 1 shows a 
simple sketch of the physical system  under consideration.

\subsection{Localized states in ballistic transport}

We have obtained the linear conductance ${\cal G}$ as a function of
the electron energy (Fermi energy) discretizing the $xy$ plane in a
uniform grid and using the quantum-transmitting-boundary algorithm \cite{len90}. 
This 
method transforms the Schr\"odinger equation into a complex linear system
whose dimension is twice the number of grid points, due to the two spin 
components. In spite of the typically large dimension, the system is highly sparse
and it can be solved efficiently with standard numerical routines. 
This approach provides the exact
wave function for the ballistic transport problem from which one 
obtains the transmission amplitudes and the wire's linear conductance.

Figure 1 shows a typical conductance curve for specific values of 
$\alpha$ and $\ell$. The usual staircase conductance is modified 
because of the Rashba dot in two main aspects. First the initial
part of the step is smoothed, a usual
quantum behavior of scattering through potential wells. The second
conspicuous feature is the existence of a pronounced conductance dip
near the end of the conductance plateau. It was already mentioned that 
the existence of conductance dips is normally due to quasibound states.
To proof it in the present context we show in Fig.\ 2 the wave function
for two different energies, $E_a$ and $E_b$, where $E_a$ is chosen 
to lie in the  smooth conductance region while $E_b$ corresponds 
precisely to the dip position. Indeed, at energy $E_b$ the wave function 
strongly resonates within the Rashba dot, being so amplified that it 
greatly exceeds the values in the leads. This explicitly shows the dramatic 
influence of the quasibound state. A systematic analysis of the dip
positions for varying $\alpha$'s and $\ell$'s was performed in 
Ref.\ \cite{san06}.

\begin{figure}[t]
\begin{minipage}{18pc}
\centerline{\includegraphics[width=16.5pc]{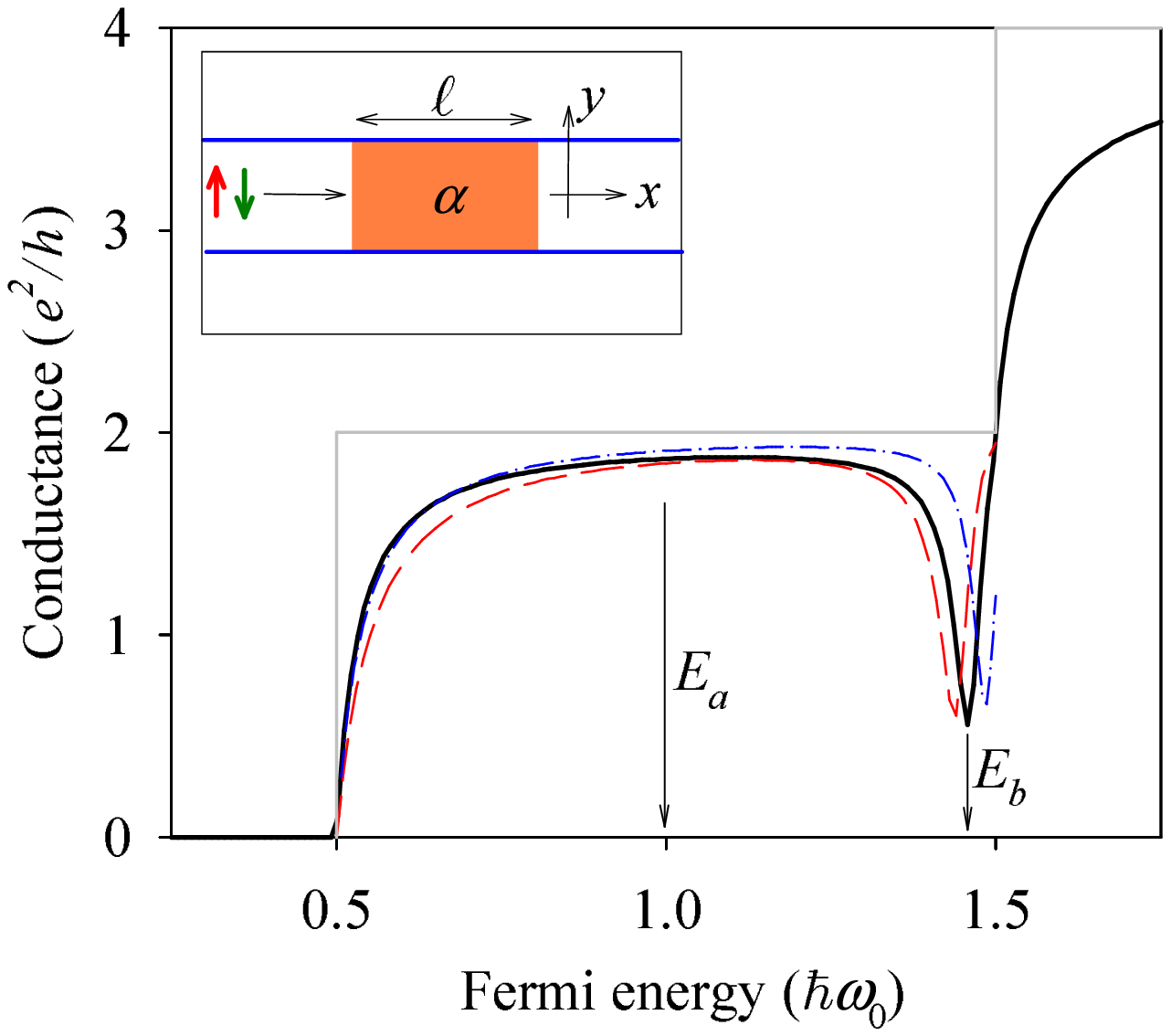}}
\caption{\label{fig1} 
(Color online)
Conductance for a Rashba dot of $\ell=\ell_0$ and $\alpha_0=0.75\hbar\omega_0\ell_0$
in the exact (solid), numerical CCM (dashed) and analytical CCM (dash-dotted) 
calculations.
The solid gray line shows the case without Rashba dot and the
inset sketches the physical system.}
\end{minipage}\hspace{2pc}%
\begin{minipage}{18pc}
\centerline{\includegraphics[width=15.9pc]{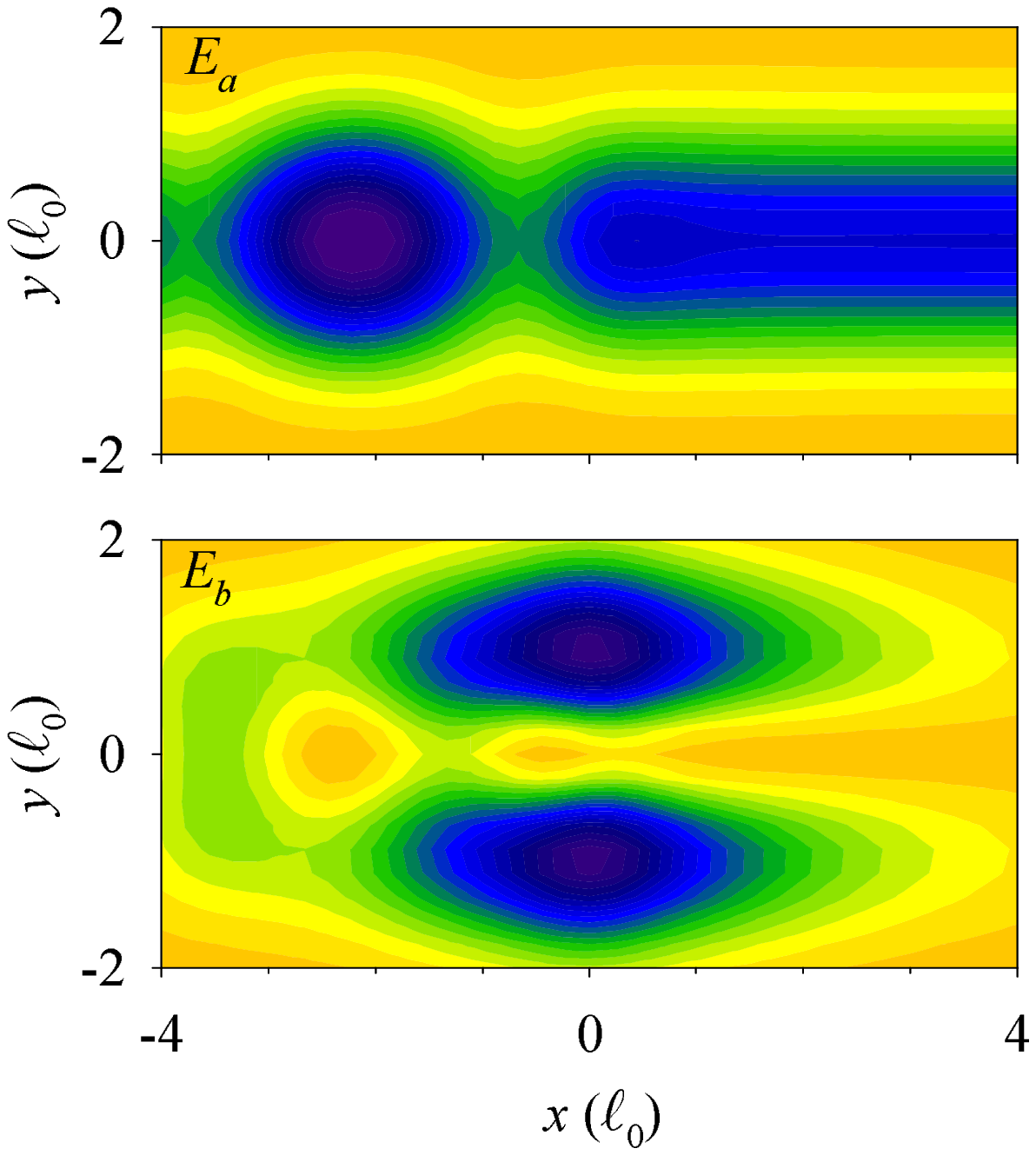}}
\caption{\label{fig2} 
(Color online)
Probability densities for the energies $E_a$ (upper)
and $E_b$ (lower) indicated in Fig.\ 1. Darker means higher probability.}
\end{minipage} 
\end{figure}

\section{The coupled channel model}

\subsection{Channel equations}

The physics behind the results shown in the preceding section is more
easily analyzed within the CCM. Assume the following expansion of the 
electron wave function $\Psi(x,y,\eta)$, with $\eta=\uparrow,\downarrow$, 
in the transverse oscillator modes $\phi_n(y)$, with energies 
$\varepsilon_n$, and spin eigenstates $\chi_\pm(\eta)$  
of $\sigma_y$,
\begin{equation}
\label{eq3}
\Psi(x,y,\eta) = \sum_{n,s=\pm}{\psi_{ns}(x)\phi_n(y)\chi_s(\eta)}\; .
\end{equation}
Substituting Eq.\ (\ref{eq3}) into the Schr\"odinger equation $H\Psi=E\Psi$
and projecting onto the functions $\phi_n\chi_s$ we obtain the CCM equations 
for the different channel amplitudes $\psi_{ns}(x)$. We shall restrict
to the first conductance plateau, i.e., to 
electron energies fulfilling $\varepsilon_1 < E < \varepsilon_2$
and also truncate the expansion in Eq.\ (\ref{eq3}) to the first
two transverse modes $n=1,2$. The selection rules stemming from the
Rashba interaction require spin flip and also $\Delta n=\pm 1$. Therefore
we obtain two independent two-band models, with $\psi_{1+}$ coupled to 
$\psi_{2-}$ and $\psi_{1-}$ to $\psi_{2+}$. Actually, the two models are 
equivalent and we shall focus for simplicity on the $\psi_{1+}$--$\psi_{2-}$ 
system.

The CCM equations for the $\psi_{1+},\psi_{2-}$ amplitudes are greatly 
simplified by means of the following gauge transformation
$\psi_{1+,2-}\rightarrow\psi_{1+,2-}\exp(\pm i \int^x{k_R(x') dx'})$,
where we have defined $k_R(x)=m\alpha(x)/\hbar^2$. The resulting 
CCM system then reads
\begin{eqnarray}
\label{eq4a}
\left[
\frac{p_x^2}{2m}-{\hbar^2k_R(x)^2\over 2m}-E+\varepsilon_1
\right] \psi_{1+}(x) &=& V_{12}(x) \psi_{2-}(x)\; ,\\
\label{eq4b}
\left[
\frac{p_x^2}{2m}-{\hbar^2k_R(x)^2\over 2m}-E+\varepsilon_2
\right] \psi_{2-}(x) &=& V_{21}(x) \psi_{1+}(x)\; .
\end{eqnarray}
The left hand sides of Eqs.\ (\ref{eq4a}) and (\ref{eq4b}) constitute the channel 
background problems. They are 1D Schr\"odinger equations 
for a potential well 
$-{\hbar^2 k_R(x)^2/2m}$ with the only difference that 
the energy is positive (negative) for $\psi_{1+}$
($\psi_{2-}$). This clearly shows the propagating and evanescent character
of the two channels, respectively. The right-hand-sides contain the 
channel couplings, with the mixing potentials 
\begin{equation}
\label{eq5}
V_{12}(x) = V_{21}^*(x) = \frac{i}{\hbar} \langle \phi_1|p_y|\phi_2\rangle
\alpha(x) e^{2i\int^x{k_R(x')dx'}}\; .
\end{equation}

\begin{figure}[t]
\includegraphics[width=18pc]{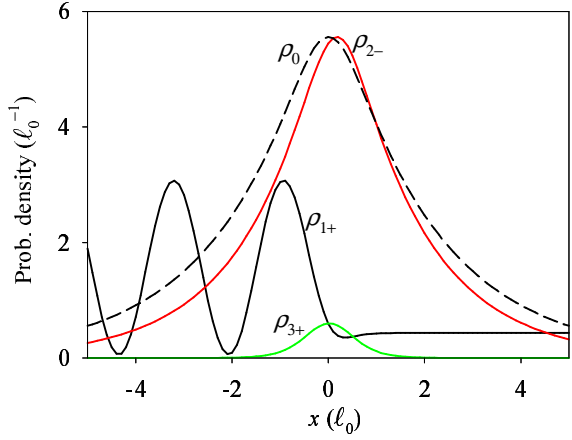}\hspace{2pc}%
\begin{minipage}[b]{18pc}\caption{\label{fig3}
(Color online)
Channel probability densities within CCM, defined as $\rho_{ns}(x)=|\psi_{ns}(x)|^2$,
for the energy $E_b$ of Fig.\ \ref{fig1}. 
The probability density corresponding to the background bound 
state $\rho_0(x)=|\phi_0(x)|^2$ is also shown.
$\rho_0(x)$ has been arbitrarily rescaled in order to reproduce the peak height of 
$\rho_{2-}(x)$ for better comparison.
}
\end{minipage}
\end{figure}

\subsection{Ansatz solution}

The evanescent character of $\psi_{2-}$ in Eq.\ (\ref{eq4b}) motivates
the ansatz $\psi_{2-}(x)\propto \phi_0(x)$, where $\phi_0$
is the background bound state fulfilling
$\left( p_x^2/ 2m - \hbar^2 k_R^2/2m-\varepsilon_0\right) \phi_0 = 0$, with 
$\varepsilon_0<0$.
Since $E>\varepsilon_1$, Eq.~(\ref{eq4a}) describes a 
1D scattering process with a source term given by 
$V_{12}\psi_{2-}$ and asymptotic wave vector 
$k=\sqrt{2m(E-\varepsilon_1)/\hbar^2}$.
Using the Green function $G$, given in terms of the 
background asymptotic states $\varphi_r$ and $\varphi_l$ behaving
as $te^{\pm ikx}$ for $x\to\pm\infty$, respectively, 
we find the total
transmission,
\begin{equation}\label{eq_t}
T_+(E)=|t|^2\frac{(E-\varepsilon_2-\varepsilon_0-\Delta+\delta)^2
+(\gamma-\Gamma)^2}{(E-\varepsilon_2-\varepsilon_0-\Delta)^2+\Gamma^2}
\equiv 
|t|^2\frac{|\epsilon+q|^2}
{\epsilon^2+1}
\; .
\end{equation}
We have defined 
$\Delta+i\Gamma \equiv \langle\phi_0 |V_{21}GV_{12}|\phi_0 \rangle$, 
$\delta+i\gamma \equiv \frac{m}{i\hbar^2kt}\langle\varphi_l^*|V_{12}|\phi_0\rangle
\langle\phi_0|V_{21}|\varphi_r\rangle$
and the last expression in Eq.\ (\ref{eq_t}) corresponds to the generalized Fano lineshape,
with $\epsilon=(E-\varepsilon_2-\tilde{\varepsilon}_0)/\Gamma$ 
and $q=\delta/\Gamma+i(\gamma/\Gamma-1)$. In general, the Rashba dot yields complex
$q$'s and therefore the factor $|\epsilon+q|$ in Eq.\ (\ref{eq_t}) can not exactly 
vanish at any $E$,
thus hindering the formation of zero transmission dips. For some particular
values of $\alpha$ and $\ell$, however, we showed in Ref.\ \cite{san06} that 
there are {\em accidental} cases 
where the conductance dip is compatible with a zero value within numerical precision.
As shown by the dash-dotted line of Fig.\ \ref{fig1}, the ansatz solution satisfactorily 
reproduces
the conductance dip although, of course, there are some minor discrepancies with the 
exact solution of the preceding section (solid line).

\subsection{Numerical solution}
The exact solution of the CCM equations can also be obtained numerically, avoiding the ansatz, 
using a 1D formulation of the 
quantum-transmitting-boundary algorithm. Dashed line in Fig.\ \ref{fig1} corresponds
to the conductance obtained from this numerical CCM approach while  
Fig.\ \ref{fig3} shows the different channel amplitudes 
(including also the $\psi_{3+}$ channel), as well as the bound 
state $\phi_0$, for the dip energy $E_b$ of Fig.\ \ref{fig1}.
The probability density for channel $\psi_{1+}$ again proves the propagating character
of this channel although transmission is rather low at this energy. 
There is good qualitative agreement of the probability densities $\rho_{2-}$ and $\rho_0$, 
explicitly proving that the ansatz is indeed a reasonable assumption 
for the evanescent channels.
The numerical solution has been obtained including also channel $\psi_{3+}$, however, 
the smallness of $\rho_{3+}$ with respect to the other channel densities
supports the truncation to the lowest two modes of the preceding subsection.

\section{Conclusions}

We have theoretically investigated the influence of a localized Rashba 
scattering center (a Rashba dot)
on the linear conductance of a quantum wire. The Rashba dot sustains quasi-bound states that 
interfere with the direct transmission along the wire and lead to Fano-resonance profiles
in the energy dependence of the linear conductance. 
This Fano-Rashba effect has been elucidated
with numerical calculations using a grid discretization of the 2D Schr\"odinger equation. 
We also analyzed a coupled-channel model, finding analytical expressions within 
the ansatz approach
that clearly show the appearance of the Fano function for the transmission.
Finally, the validity of the ansatz solution and the truncation to the lowest two modes
has been assessed by finding the numerical solution of the 
multimode coupled-channel-model equations.

\ack
We acknowledge R. L\'opez for valuable discussions.
This work was supported by the Grant No.\ FIS2005-02796
(MEC) and the ``Ram\'on y Cajal'' program.

\end{document}